# A common framework for single-molecule localization using sequential structured illumination

Luciano A. Masullo[1,2], Lucía F. Lopez[1], Fernando D. Stefani[1,2*]

[1] Centro de Investigaciones en Bionanociencias (CIBION), Consejo Nacional de Investigaciones Científicas y Técnicas (CONICET), Godoy Cruz 2390, C1425FQD, Ciudad Autónoma de Buenos Aires, Argentina

[2] Departamento de Física, Facultad de Ciencias Exactas y Naturales, Universidad de Buenos Aires, Güiraldes 2620, C1428EHA, Ciudad Autónoma de Buenos Aires, Argentina

*fernando.stefani@df.uba.ar

## ABSTRACT

Localization of single fluorescent molecules is key for physicochemical and biophysical measurements such as single-molecule tracking and super-resolution imaging by single-molecule localization microscopy (SMLM). Recently a series of methods have been developed in which the localization precision is enhanced by interrogating the molecular position with a sequence of spatially modulated patterns of light. Among them, the MINFLUX technique outstands for achieving a ~10-fold improvement compared to wide-field camera-based single-molecule localization, reaching ~1 − 2 nm localization precision at moderate photon counts. Here, we present a common mathematical framework for this type of measurement that allows a fair comparison between reported methods and facilitates the design and evaluation of new methods. With it, we benchmark all reported methods for single-molecule localization using sequential structured illumination, including long-established methods such as orbital tracking, along with two new proposed methods: orbital tracking and raster scanning with a minimum of intensity.

# INTRODUCTION

Since it became technically possible, localization of single fluorescent molecules has been key to obtain information on biological processes beyond ensemble averages. For instance, single-molecule tracking measurements provide unique insight into molecular trajectories that would otherwise be hidden in the average behavior of an ensemble of unsynchronized molecules[1–5]. Another important application of single-molecule localization is single-molecule localization microscopy (SMLM) methods. In SMLM, single-molecule localization is combined with single-molecule blinking in order to determine the positions of a multitude of molecules in a sample. In this way, super-resolved fluorescence images can be reconstructed where the spatial resolution is ultimately given by the localization precision[6,7].

The performance of single-molecule tracking and SMLM is limited by the photostability of the fluorophores[7–9]. Most commonly, single-molecule localization is performed using uniform illumination, and the position of the molecule is determined from a fit to its image recorded with a photodetector array such as an EM-CCD or a CMOS camera. With this approach, the lateral localization precision of organic fluorophores under biologically compatible conditions lies typically in the range of $10 - 50$ nm. Recently, aiming to attain higher localization precisions with the available photon budget, a series of methods have been developed where single emitters are interrogated with a sequence of spatially modulated patterns of light. This new trend of measurements was opened by the publication of MINFLUX[10], achieving a ~10-fold improvement compared to wide-field camera-based single-molecule localization, reaching $\sim 1 - 2$ nm localization precision at moderate photon counts. Since then, MINFLUX has been demonstrated in model systems (DNA-origami structures), fixed and living cells, and it was recently extended to three dimensions[11]. Also, other methods of this kind have been reported, such as ROSE[12], SIMFLUX[13], MINSTED[14], and MODLOC[15]. This type of single-molecule localization has been recently reviewed[16].

On the other hand, around twenty years ago, before the advent of SMLM, a method to track the motion of particles or single fluorescent molecules in 2D called Orbital Tracking (OT) was theoretically proposed[17] and later implemented experimentally in a multitude of situations including 3D tracking and combinations with fluorescence correlation spectroscopy[18–22]. In OT, the fluorescence signal from a single particle or molecule is registered for a number of positions along a circular trajectory of a focused laser beam around the target molecule or particle. Other methods of single-molecule tracking based on multiple exposures of displaced focused beams have also been reported, such as the four-focus single-particle position determination[23,24]. To the best of our knowledge, these localization techniques developed for tracking have not been combined with single-molecule blinking in order to obtain super-resolved images.

At first sight, due to the differences in the structure of the excitation light, instrumentation, measurement protocols, and data analysis methods, each of these methods of single-molecule localization may appear unique. Here, we show how these techniques can be regarded as special cases of a common concept of single-molecule localization using sequences of excitations with spatially structured light. We present a common analytical framework for this type of single-molecule localization and use it to i) perform a fair benchmarking between methods and ii) identify new single-molecule localization methods that bring together the strengths of the available techniques.

**METHODS**

A COMMON FRAMEWORK FOR SINGLE-MOLECULE LOCALIZATION USING SEQUENTIAL STRUCTURED ILLUMINATION

Figure 1a shows schematically the essential components of single-molecule localization by sequential structured illumination. A spatially structured excitation field $I(\boldsymbol{r})$ is sequentially shifted along a sequence of $K$ positions $\boldsymbol{r}_i$ ($1 \leq i \leq K$). In this paper, we will deal with the two-dimensional (2D) localization problem. Naturally, the formalism can be easily reduced to 1D localization or extended to 3D localization. In 2D, the $K$ positions $\boldsymbol{r}_i$ may be arbitrary within the plane of interest but must not be in line to avoid obvious localization ambiguities. We will call the sequence of $I(\boldsymbol{r} - \boldsymbol{r}_i)$ the "excitation pattern", and $\boldsymbol{r}_E$ the position of the emitter. For each $I(\boldsymbol{r} - \boldsymbol{r}_i)$, the emitter is exposed to a specific local intensity $I(\boldsymbol{r}_E - \boldsymbol{r}_i)$ and emits fluorescence with a certain intensity, which in turn corresponds to an expected value of detected photon counts ($\lambda_i$) during a given integration time. The measured fluorescence photon counts are denoted by $n_i$, which are assumed to be Poisson distributed with average $\lambda_i$. The latter is an excellent approximation for modern avalanche photodiodes (APD) with neglectable dark counts and readout noise. The position of the emitter is determined from the sequence of intensity measurements $\boldsymbol{n} = [n_1, n_2, ..., n_K]$, and considering the known $I(\boldsymbol{r} - \boldsymbol{r}_i)$. The relationship between $I(\boldsymbol{r}_E - \boldsymbol{r}_i)$ and $\lambda_i$ is assumed to be linear (emission far from saturation).

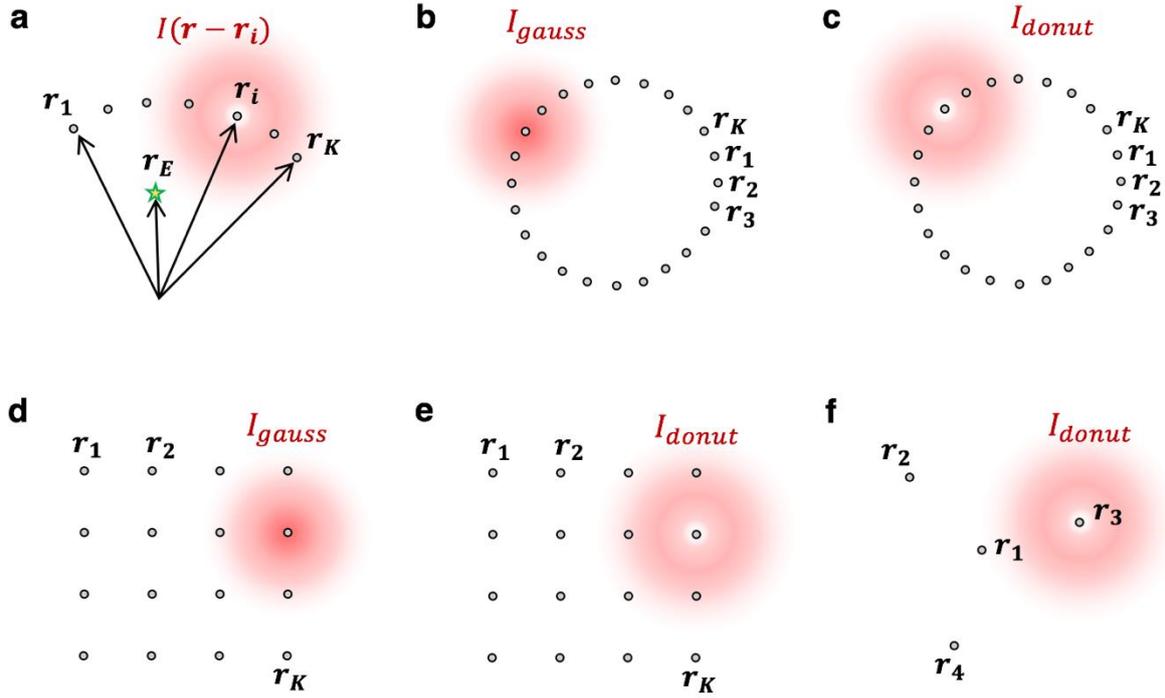

**Figure 1**. (a) Schematic showing the essential parameters of a SML-SSI measurement in 2D. $r$ defines the position in the plane of interest. $I(r - r_i)$ is the structured excitation field located at position $r_i$. The excitation field is sequentially placed at $K$ positions $r_i$ ($1 \leq i \leq K$). At each position of the excitation field, the intensity of an emitter placed at $r_E$ is registered. (b-f) Example configurations of SML-SSI measurements using a maximum ($I_{gauss}$) or a minimum ($I_{donut}$) of light.

Any method of single-molecule localization using sequential structured illumination can be fully described by the set of $I(r - r_i)$, which in turn is defined by the spatial structure of the excitation field $I(r)$ and the sequence of positions of the exposures $r_i$.

We will deal with methods using focused laser beams, which can be classified into two categories depending on whether the focus has a central maximum or a central minimum (ideally a zero) of intensity. For our analysis, focused excitation fields with a central maximum will be described with a Gaussian function:

$$I_{Gauss}(r) = A_0 e^{-4\ln 2 \frac{r^2}{FWHM^2}} \qquad (1)$$

and excitation fields with a central zero, here called donut-shaped foci, will be described as:

$$I_{donut}(r) = A_0 4e \ln 2 \frac{r^2}{FWHM^2} e^{-4 \ln 2 \frac{r^2}{FWHM^2}} \qquad (2)$$

While for the following calculations we will use the idealized $I_{Gauss}(r)$ and $I_{donut}(r)$, we note that the analysis can be performed with any other shape of $I(r)$, particularly with functions describing more accurately experimentally determined illumination patterns. Here, we will treat $I(r)$ as a known function. In experiments, $I(r)$ must be determined. For this reason, SML-SSI methods usually involve two measurements: **(1)** A detailed characterization of the excitation light field $I(r)$ using bright emitters (i.e. fluorescent nanoparticles) delivering almost unlimited photon counts (i.e. $N > 10^6$, high SNR), and **(2)** the measurement with limited photon counts (i.e. $N < 10^3$, low SNR) by sequentially exciting the single emitter (i.e. organic fluorophore or fluorescent protein), whose position is unknown.

As for the sequence of excitation positions $r_i$, we will consider two types too: orbital sequences enclosing an area (as it is done in orbital tracking), and raster-scanning sequences covering an area (as it is done in raster-scanning microscopy). Varying combinations of $I_{Gauss}$, $I_{donut}$ and sequence of $r_i$ can be used to define any single-molecule localization method using sequential illumination with focused beams, including all reported methods and any new conception. For example, Figure 1b shows schematically the combination used for classical orbital tracking (OT)[17,18,25], namely, $I_{Gauss}$ excitation sequentially shifted over $K$ positions along a circle. In practice, optimum performance in OT is achieved with a radius of the circle close to half the full-width at half-maximum (FWHM) of $I_{gauss}$ [18,25,26]. The number of exposures $K$ may vary from a few up to a quasi-continuum intensity register. The so-called Single-Molecule Confocal Laser Tracking (SMCT)[22] can be regarded as a special case of orbital tracking with $K = 6$. MINSTED[14] is, in essence, another expression of OT that achieves higher localization precision by using an effectively smaller excitation field, produced by the combination of a normal excitation beam and a donut-shaped depletion beam, just as in STED microscopy[27,28]. Hence, the excitation field of MINSTED can be described by $I_{Gauss}(r)$ with a FWHM below the diffraction limit. Alternatively, OT could be performed with $I_{donut}(r)$, as schematically shown in Figure 1c. We will call this method OTMIN. So far, it has not been proposed or implemented.

The sequence of $r_i$, can also be organized in a raster to cover an area, as shown in Figure 1d for $I_{Gauss}(r)$. This configuration, here denoted RASTMAX, has been recently applied in a conventional confocal microscope[29]. Under this framework, a new method where $I_{donut}(r)$ is raster scanned over a rectangular area can be easily envisaged, as schematically shown in Figure 1e; we will call this new scheme RASTMIN. Finally, Figure 1f shows the scheme of 2D MINFLUX where $I_{donut}(r)$ is shifted over four positions: a central exposure and three more forming an equilateral triangle around the central position[30,31]. 2D MINFLUX can be classified as a raster-scanning method because the excitation pattern used is the minimum needed to cover an area.

POSITION ESTIMATION AND PRECISION

Estimating the molecular position from the intensity measurements $\boldsymbol{n} = [n_1, n_2, ..., n_K]$ and $I(\boldsymbol{r} - \boldsymbol{r}_i)$ can be done in innumerable ways, and many have been implemented in the various methods cited above. For example, in orbital tracking, the position of the emitter has been estimated by analyzing quasi-continuum intensity signals by Fourier analysis[18] or by triangulation of discrete intensity signals[22]. In MINFLUX[30,31], or the four-focus single-particle localization[23], the position of the emitter is obtained using a maximum likelihood estimator with four intensity measurements. Other methods such as MINSTED[14] use other ad-hoc analysis functions and routines.

Ideally, the position estimator must be unbiased and accurate. Independently of the estimator used, using the Fisher information matrix, a theoretical maximum accuracy for an unbiased position estimator can be calculated in the form of a theoretical lower bound for the variance of the estimator, the so-called Cramér-Rao bound (CRB)[32]. Here, we will use the maximum likelihood estimator to determine the emitter position from $\boldsymbol{n}$ and $I(\boldsymbol{r} - \boldsymbol{r}_i)$, as it is by far the most widely used approach in statistical estimation due to its well-established performance; it is in general consistent and asymptotically attains the CRB[32]. Studies about deconvolution in microscopy showed that the MLE is more accurate than least squares based algorithms especially for quantum-limited data, i.e. Poisson distributed data with low signal levels[33,34]. More recently the performance of the MLE for single-molecule localization has been validated in camera-based approaches[35–37] as well as in sequential structured illumination approaches such as the 3D four-focus localization[23] or MINFLUX[10].

The likelihood function $\mathcal{L}$ for the emitter position can be expressed as:

$$\mathcal{L}(\boldsymbol{r}_E|\boldsymbol{n}) = \frac{N!}{\prod_{i=1}^{K} n_i!} \prod_{i=1}^{K} p_i(\boldsymbol{r}_E)^{n_i} \qquad (3)$$

where $N = \sum_{i=1}^{K} n_i$ is the total number of detected photons, and $p_i(\boldsymbol{r_E})$ is the multinomial parameter for each exposure:

$$p_i(\boldsymbol{r_E}) = \frac{I(\boldsymbol{r_E}-\boldsymbol{r_i})}{\sum_{j=1}^{K} I(\boldsymbol{r_E}-\boldsymbol{r_j})} \tag{4}$$

defined as the ratio between the intensity of the excitation field at the fluorophore position for the current exposure and the sum of all the exposure intensities. In the presence of background, defined by the signal-to-background ratio (SBR) eq. (4) becomes:

$$p_i(\boldsymbol{r_E}) = \frac{SBR(\boldsymbol{r_E})}{SBR(\boldsymbol{r_E})+1} \frac{I(\boldsymbol{r_E}-\boldsymbol{r_i})}{\sum_{j=1}^{K} I(\boldsymbol{r_E}-\boldsymbol{r_j})} + \frac{1}{SBR(\boldsymbol{r_E})+1} \frac{1}{K} \tag{5}$$

where

$$SBR(\boldsymbol{r_E}) = \frac{\sum_{j=1}^{K} I(\boldsymbol{r_E}-\boldsymbol{r_j})}{\sum_{j=1}^{K} I_b(\boldsymbol{r_E})} = \frac{\sum_{j=1}^{K} I(\boldsymbol{r_E}-\boldsymbol{r_j})}{K I_b} \tag{6}$$

Here, we have assumed that the background contribution is equal for all exposures and does not depend on the position of the emitter. A detailed derivation of eq. (5) is described in Supplementary Section 1. $SBR(\boldsymbol{r_E})$ can be calculated from an assumption (or experimental determination) of $SBR$ at the center of the excitation pattern $SBR(\boldsymbol{0})$, as

$$SBR(\boldsymbol{r_E}) = SBR(\boldsymbol{0}) \frac{\sum_{j=1}^{K} I(\boldsymbol{r_E}-\boldsymbol{r_j})}{\sum_{j=1}^{K} I(\boldsymbol{0}-\boldsymbol{r_j})} \tag{7}$$

In the following, we will use $SBR(\boldsymbol{0}) \equiv SBR$ as a scalar parameter for the benchmarking of the different methods.

For the MLE, it is practical to use the log-likelihood function $l(\boldsymbol{r_E}|\boldsymbol{n}) = \ln(\mathcal{L}(\boldsymbol{r_E}|\boldsymbol{n}))$:

$$l(r_E|n) = \sum_{i=0}^{K} \ln(p_i(r_E)) n_i \tag{8}$$

since we are interested in finding the value of $r_E$ that maximizes the function. In eq. (8), all additive constants have been omitted because they are irrelevant for the maximum likelihood estimation of the emitter position, which is computed as follows:

$$\widehat{r_E}^{MLE} = \arg\max (l(r_E|n)) \tag{9}$$

In general, single-molecule localization by sequential structured illumination delivers high precision position estimations only for molecules in the vicinity of the excitation pattern. Thus, extra, prior, lower-precision information about the emitter position is necessary to place the excitation pattern in such a way that the emitter position can be estimated with high precision. The likelihood function can be modified to include this *prior* as follows:

$$\mathcal{L}(r_E|n) = \frac{N!}{\prod_{i=1}^{K} n_i!} \prod_{i=1}^{K} p_i(r_E)^{n_i} f(r_E) \tag{10}$$

Where the function $f(r_E)$ includes the prior information about the emitter position. The log-likelihood function then becomes:

$$l(r_E|n) = \sum_{i=1}^{K} n_i \ln p_i(r_E|n) + \ln f(r_E) \tag{11}$$

Where, again, all the constant terms have been dropped since we are only interested in the maximum of the $l(r_E|n)$ function. We note that $f$ may depend on an independent set of photon counts used to determine the molecule position with low precision.

For the 2D problem, $r_E = (x, y)$ and the Fisher information matrix takes the form:

$$\mathcal{I}(r_E) = -E\left(\begin{bmatrix} \frac{\partial^2 l(r_E|n)}{\partial x^2} & \frac{\partial^2 l(r_E|n)}{\partial x \partial y} \\ \frac{\partial^2 l(r_E|n)}{\partial y \partial x} & \frac{\partial^2 l(r_E|n)}{\partial y^2} \end{bmatrix}\right) \tag{12}$$

which using eq. (11) can be expressed as:

$$\mathcal{I}(\boldsymbol{r_E}) = \mathcal{I}_{SML-SSI} + \mathcal{I}_{prior} = N \sum_{i=1}^{K} \frac{1}{p_i} \begin{bmatrix} \left(\frac{\partial p_i}{\partial x}\right)^2 & \frac{\partial p_i}{\partial x}\frac{\partial p_i}{\partial y} \\ \frac{\partial p_i}{\partial y}\frac{\partial p_i}{\partial x} & \left(\frac{\partial p_i}{\partial y}\right)^2 \end{bmatrix} - \begin{bmatrix} \frac{\partial^2 \ln f}{\partial x^2} & \frac{\partial^2 \ln f}{\partial x \partial y} \\ \frac{\partial^2 \ln f}{\partial y \partial x} & \frac{\partial^2 \ln f}{\partial y^2} \end{bmatrix} \quad (13)$$

Finally, the lower bound for the covariance matrix of the estimated emitter position as a function of the real emitter position, $\Sigma_{cov}(\boldsymbol{r_E})$, can be obtained from the Cramér-Rao inequality:

$$\Sigma_{cov}(\boldsymbol{r_E}) \geq \Sigma_{CRB}(\boldsymbol{r_E}) = \mathcal{I}(\boldsymbol{r_E})^{-1} \quad (14)$$

For simplicity, we will take the arithmetic mean of the eigenvalues of $\mathcal{I}(\boldsymbol{r_E})^{-1}$ as a measure of the average maximum precision:

$$\sigma_{CRB}(\boldsymbol{r_E}) = \sqrt{\frac{1}{2} tr\left[\Sigma_{CRB}(\boldsymbol{r_E})\right]} = \sqrt{\frac{1}{2\, det[\mathcal{I}(\boldsymbol{r_E})]} tr\left[\mathcal{I}(\boldsymbol{r_E})\right]} \quad (15)$$

In general, $f(\boldsymbol{r_E})$ reduces the uncertainty in the position estimation. To visualize this, it can be considered that any *prior* can be expressed, at least approximately, as a Gaussian function or similar centered at the estimated position, whose logarithm has a second derivative that is always negative.

The implementation of this mathematical framework, i.e. all functions and scripts used in this work, is written in Python and is fully open-source. It can be found at https://github.com/lumasullo/sml-ssi and https://github.com/stefani-lab/sml-ssi. All calculations and simulations can be easily reproduced following the instructions in the repositories.

**RESULTS AND DISCUSSION**

BENCHMARKING DIFFERENT METHODS

Next, we benchmark the theoretical performance of different orbital and raster scanning methods, including reported techniques and new proposals. For each method, we show an exemplary 2D map of $\sigma_{CRB}$ for a set of realistic experimental parameters ($N = 500$ detected photons, $SBR = 5$) and then compute the average $\sigma_{CRB}$ ($\bar{\sigma}_{CRB}$) within a circular field-of-view ($FOV$) concentric with the excitation pattern.

The size of the excitation pattern is a relevant parameter for all methods. Here, we will parametrize it by $L$, the diameter of the orbit or the diagonal of the raster, for orbital or scanning methods, respectively. For a $FOV$ with a diameter of $0.75\ L$, which is a suitable localization region for all methods, we evaluate $\bar{\sigma}_{CRB}$ as a function of $N$ and $SBR$. In all cases, we used a Gaussian *prior* $f(\mathbf{r_E})$ that represents a rough previous localization of the emitter ($\sigma_{prior} = 50$ nm), which is a common step of all real-life experiments of this kind. The cost in photon budget of this prior information is not analyzed as it would be the same for all the methods; it would lay in the $N = 50 - 100$ range, depending on the procedure used.

ORBITAL METHODS

We first analyze orbital methods using $I_{gauss}$ excitation. We note that, theoretically, the localization precision using Gaussian beams increases indefinitely with $L$. However, in practice, the drop in $SBR$ leads to a compromise value of the orbit roughly equal to the $FWHM$[18,26]. Therefore, all orbital methods using a Gaussian beam will be studied for $L = FWHM$.

Figure 2a shows a map of the localization precision ($\sigma_{CRB}$) for orbital tracking (OT) with $L = FWHM = 300$ nm, $K = 100$, $N = 500$, and $SBR = 5$. The performance is approximately flat in areas up to $\sim L^2$. This behavior is also evident in the curves of $\bar{\sigma}_{CRB}$ vs. size of the $FOV$ for OT ($L = 300$ nm) and MINSTED ($L = 100$ nm and $L = 50$ nm) in Figure 2b. For the case of $L = 50$ nm, it can be observed that the localization uncertainty increases up to $20 - 30$ nm for $FOV > 5L$. A similar behavior is observed for all orbital tracking implementations scaled by $L$. Also in Figure 2b, the performance of these methods is shown for $K = 6$ (dotted lines). Particularly, the diffraction-limited case ($L = 300$ nm) with $K = 6$ corresponds to the method reported as SMCT. The theoretical localization precision achieved with just 6 exposures is practically the same as with its quasi-continuous counterpart ($K = 100$).

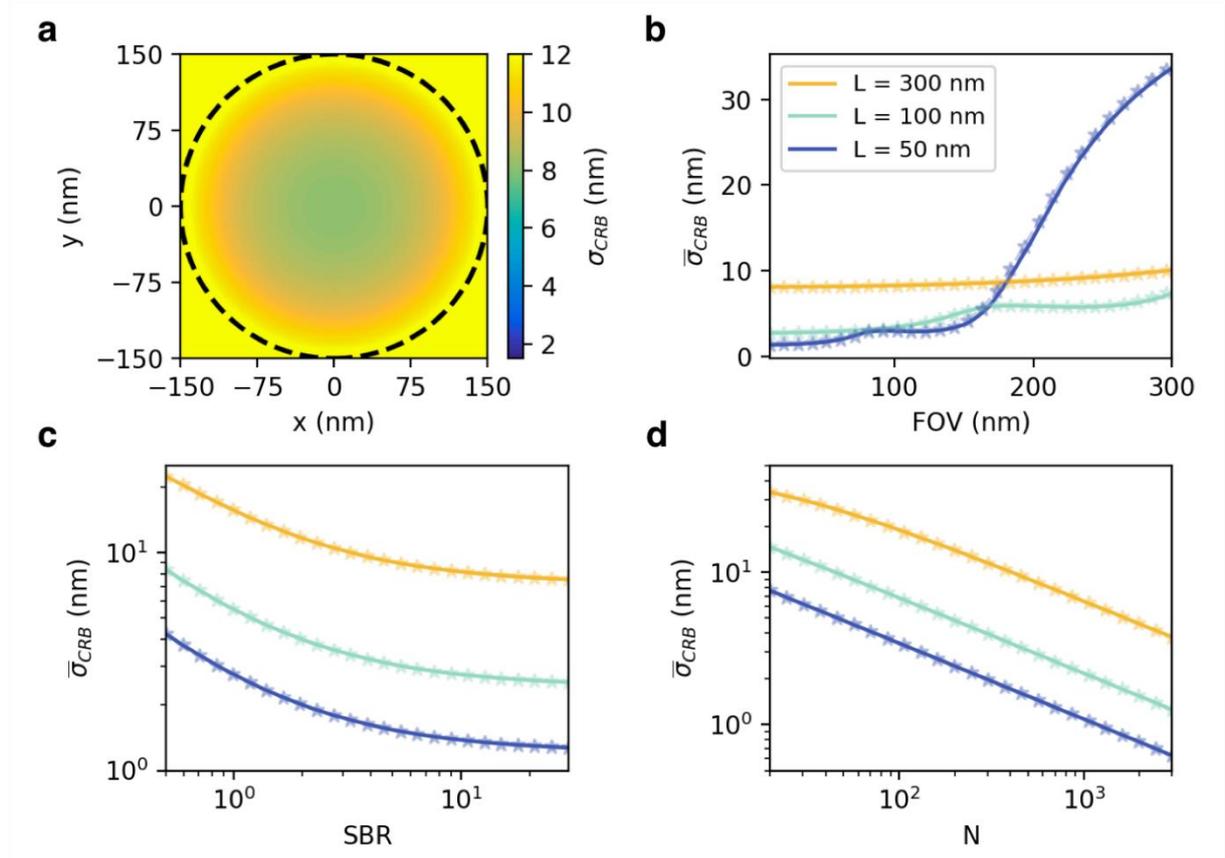

**Figure 2. Orbital Tracking, Single-Molecule Confocal Tracking, and MINSTED.** (a) Precision map $\sigma_{CRB}(x, y)$ for $L = FWHM = 300$ nm, $K = 100$. The black dotted line indicates the orbit. (b) $\bar{\sigma}_{CRB}$ as a function of the $FOV$ for $K = 100$ (solid) and $K = 6$ (stars) for three values of $L = FWHM$. (c) $\bar{\sigma}_{CRB}$ as a function of $SBR$. (d) $\bar{\sigma}_{CRB}$ as a function of $N$. Parameters: $N = 500$, $SBR = 5$ unless otherwise stated.

Figures 2c and 2d show the $\bar{\sigma}_{CRB}$ over a $FOV$ with a diameter of $0.75\,L$ as a function of $SBR$ and $N$, respectively. Both continuous (solid line) and discrete (stars) versions show almost identical behaviors and are strongly influenced by the size of $FWHM = L$, which explains the better precision achieved with MINSTED. Attaining 1-nm precision with $N = 1000 - 3000$ is only possible with $L < 100$ nm, i.e. by means of STED or any other way to achieve sub-diffraction effective excitation fields.

Next, we analyze the performance of a method featuring a minimum of intensity in the excitation beam ($I_{donut}$) and an orbital sequence of exposures. To our knowledge, such a method has not been realized experimentally. We will refer to it as Orbital Tracking with a MINimum (OTMIN).

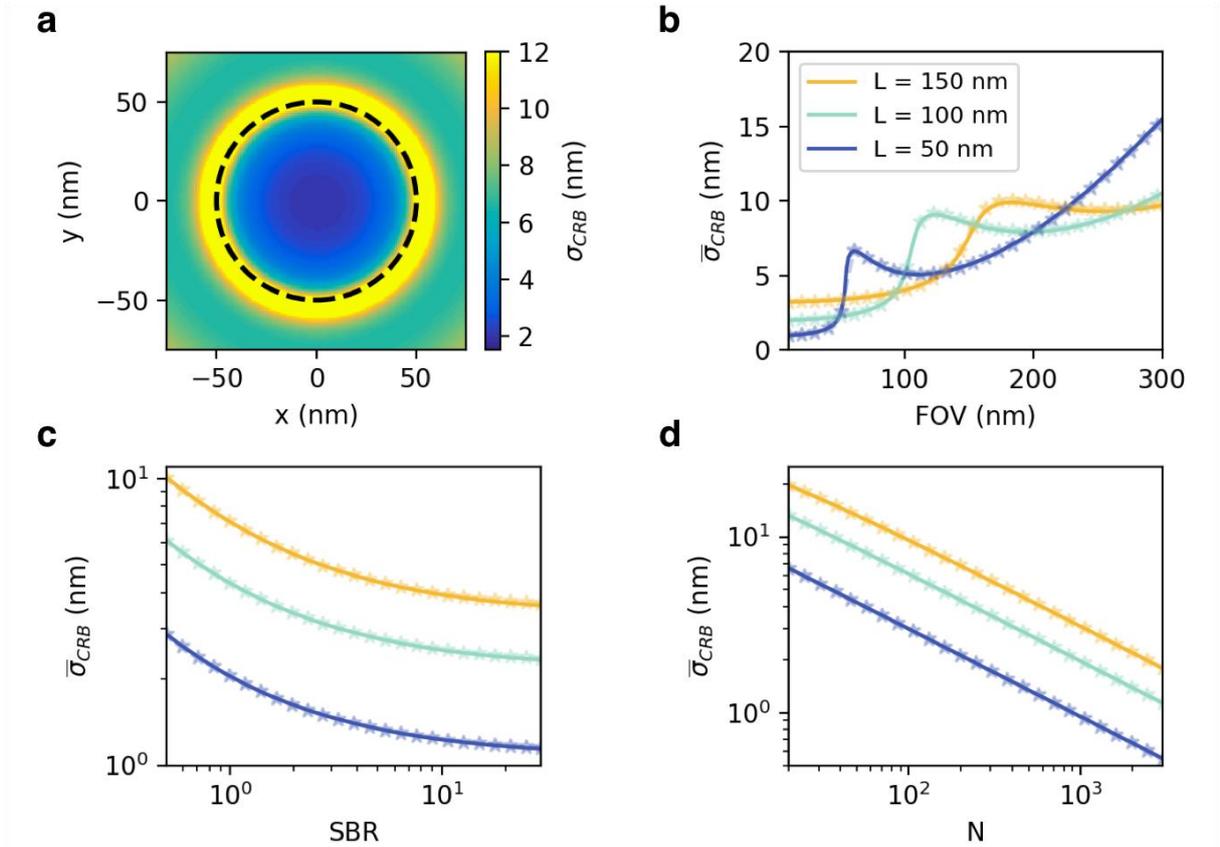

**Figure 3. Orbital Tracking with a minimum of intensity.** (a) Precision map $\sigma_{CRB}(x, y)$ for $L = 100$ nm. The black dotted line indicates the orbit. (b) $\bar{\sigma}_{CRB}$ as a function of the $FOV$. (c) $\bar{\sigma}_{CRB}$ as a function of $SBR$. (d) $\bar{\sigma}_{CRB}$ as a function of N. Parameters: $K = 100$ (solid) and $K = 6$ (stars), $FWHM = 300$ nm, $N = 500$, $SBR = 5$, unless otherwise stated.

Figure 3a shows a 2D map of $\sigma_{CRB}$ for OTMIN with $L = 100$ nm and $K = 100$. In contrast to OT, OTMIN can be performed with orbits of arbitrarily small size without the need of applying sub-diffraction techniques. OTMIN delivers accurate localizations in the inner part of the orbit. Remarkably, in the region close to the orbit the $\sigma_{CRB}$ increases rapidly. Monte-Carlo simulations confirm that the OTMIN estimator is accurate and reaches the CRB in the inner part of the area defined by the orbit (Supplementary Figure 1a) but becomes imprecise and inaccurate in the vicinity of the orbit (Supplementary Figure 1b). Nonetheless, this ill-behaved region is very narrow. Experimentally, it could be avoided by injecting information to the measurement in order to use a $FOV$ limited to the well-behaved area, e.g., periodically recentering the pattern in real-time.

Figure 3b shows curves of $\bar{\sigma}_{CRB}$ vs. size of the $FOV$ for OTMIN with $L = 50, 100$, and $150$ nm for $K = 100$ (solid) and $K = 6$ (stars). The performance of OTMIN is practically identical for $K = 100$ and $K = 6$. The best achievable localization precision of OTMIN improves with decreasing values of $L$ (for a constant $FWHM = 300$ nm of the focused beam). This increase in localization precision at the expense of limiting the $FOV$ is a common feature of all methods using a minimum of intensity. Experimentally, the ultimate limit to shrink $L$ is the decrease in $SBR$. While the FOV can have a sub-diffraction size, the illumination and detection volumes are still diffraction-limited. Thus, for a given illumination intensity, reducing $L$ to subdiffraction dimensions reduces the excitation and fluorescence emission of the emitters, but the background contribution remains constant.

In all cases, for a $FOV$ size of up to $0.75\,L$, the average localization precision of OTMIN remains remarkably high. For example, for $N = 500$ and $SBR = 5$, OTMIN reaches an average precision of $\bar{\sigma}_{CRB} < 2$ nm with $L = 100$ nm, or $\bar{\sigma}_{CRB} < 1$ nm with $L = 50$ nm (Figure 3c-d). This level of performance is only comparable to the best-reported localization precision, attained with MINFLUX. OTMIN could be of particular interest for several labs in the world that already have OT setups. Their localization precision could be increased significantly simply by adding a suitable phase-mask in the excitation path to generate a focus with a central minimum.

RASTER METHODS

MINFLUX, using just four exposures ($K = 4$) with the excitation pattern $I_{donut}$, can be regarded as the minimal expression of a raster method. Three of the exposures delimit an area that is probed with just one central exposure. MINFLUX performance has been comprehensively studied both theoretically and experimentally[10,11]. Here, we reproduce (for completeness) and expand the reported theoretical results. We note, however, that our calculations include the spatial dependency of $SBR(x, y)$ instead of using the approximation of a constant $SBR(x, y) = SBR(0, 0)$. Figure 4a shows a map of $\sigma_{CRB}$ for MINFLUX with $L = 100$ nm, $N = 500, SBR = 5, FWHM = 300$ nm. Figure 4b displays curves of $\bar{\sigma}_{CRB}$ vs size of the $FOV$ for $L = 50, 100$, and $150$ nm ($N = 500$ and $SBR = 5$), the central exposure of MINFLUX directly solves the problems of OTMIN close to the orbit border and no local maximum in uncertainty appears for $FOV \sim L$. As already reported, MINFLUX delivers the best localization precision at the center of the excitation pattern; a common feature of all these methods. For instance, with $N = 500, SBR = 5$, and $L = 100$ nm, the average precision is

$\bar{\sigma}_{CRB} = 2.7$ nm for a $FOV = 0.75\,L$, while the precision at the center of the excitation pattern is $\sigma_{CRB}(0,0) = 2.0$ nm. The localization precision of MINFLUX is the best demonstrated to date, achieving $\bar{\sigma}_{CRB} < 1$ nm for $L = 50$ nm and $N \geq 800$, $SBR \geq 5$ (Figure 4d). It should be noted that as $L$ is decreased, the precision at the center of the excitation pattern increases but $\bar{\sigma}_{CRB}(FOV)$ grows more rapidly (Figure 4b), specially outside the region defined by the excitation pattern. For example, $\bar{\sigma}_{CRB}(FOV = 200 \text{ nm}) \sim 6$ nm for $L = 100$ nm, while $\bar{\sigma}_{CRB}(FOV = 200 \text{ nm}) \sim 10$ nm for $L = 50$ nm.

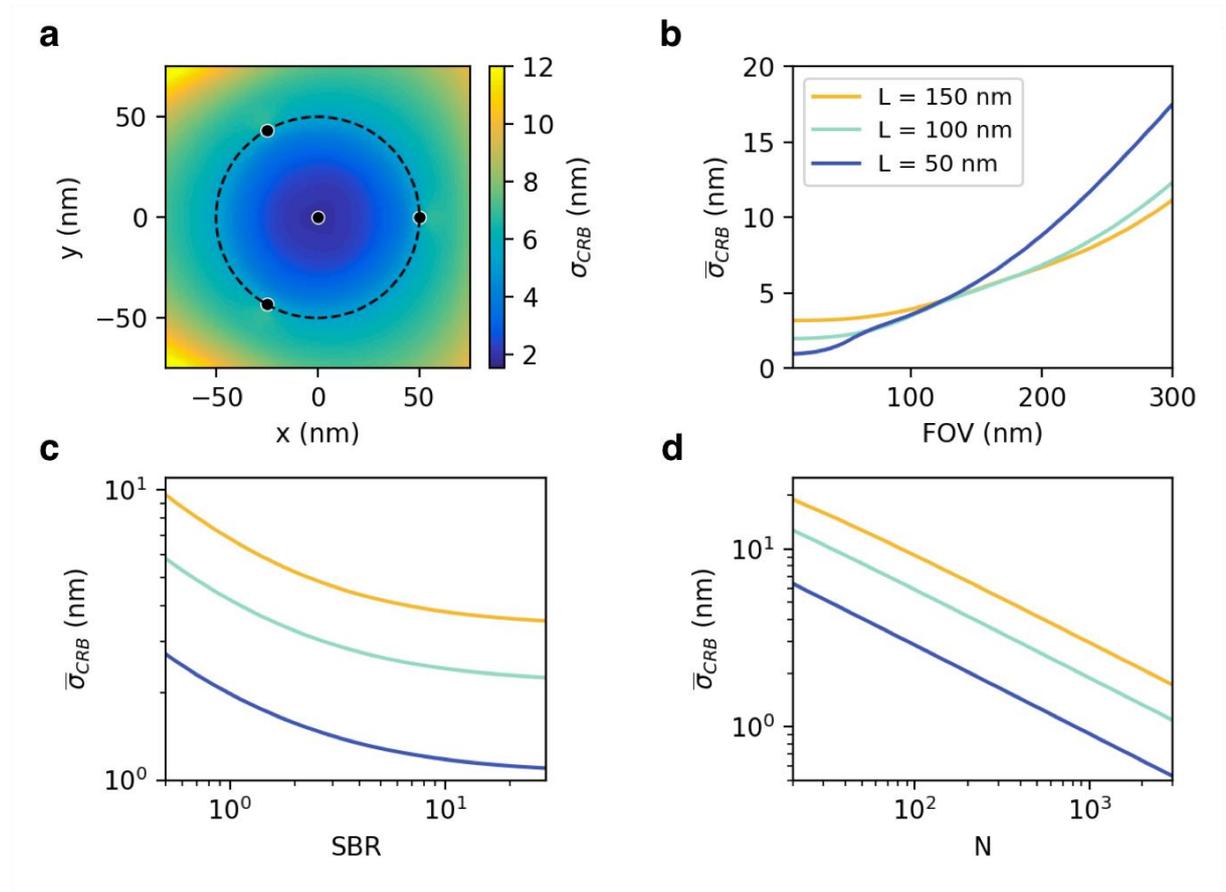

**Figure 4. MINFLUX.** (a) Precision map $\sigma_{CRB}(x,y)$ for $L = 100$ nm. Black dotted line indicates a circle of diameter $L$, black dots indicate the positions $r_i$ of the exposures. (b) $\bar{\sigma}_{CRB}$ as a function of the $FOV$. (c) $\bar{\sigma}_{CRB}$ as a function of $SBR$. (d) $\bar{\sigma}_{CRB}$ as a function of N. Parameters: $K = 100$, $FWHM = 300$ nm, $N = 500$, $SBR = 5$ unless otherwise stated.

Another method of this kind consists of using exposures of a minimum of intensity organized in a rectangular raster. To our knowledge, such a method has not been reported either theoretically or experimentally. We will refer to it as RASTer scanning with a MINimum

(RASTMIN). Figure 5a shows a 2D map of $\sigma_{CRB}(x,y)$ for RASTMIN with $L = 100$ nm, $N = 500$, and $SBR = 5$. As it happens with MINFLUX, the central exposures in RASTMIN solve the ill-behaved area problem that appears in OTMIN for $FOV \sim L$ (Figure 5b).

The performance of RASTMIN in terms of $SBR$ (Figure 5c) and $N$ (Figure 5d) is very similar to MINFLUX and OTMIN, achieving its best performance for $SBR > 5$ and reaching precisions of $\sim 1$ nm for $N = 500$ and $N = 1000$ for $L = 50$ nm and $L = 100$ nm respectively (Figure 5d).

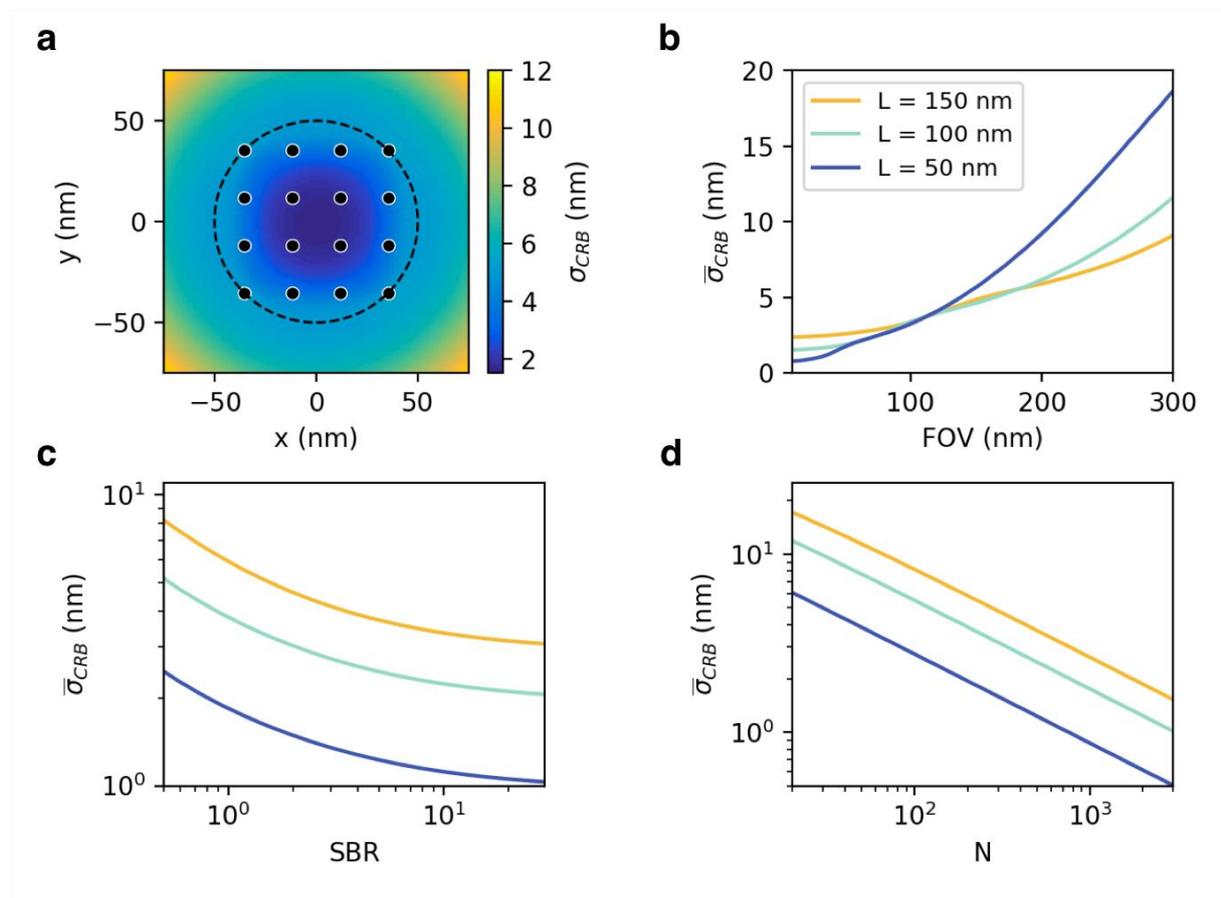

**Figure 5. Raster scanning with a minimum.** (a) Precision map $\sigma_{CRB}(x,y)$ for $L = 100$ nm. Black dotted line indicates a circle of diameter $L$, black dots indicate the positions $r_i$ of the exposures. (b) $\bar{\sigma}_{CRB}$ as a function of the $FOV$. (c) $\bar{\sigma}_{CRB}$ as a function of $SBR$. (d) $\bar{\sigma}_{CRB}$ as a function of N. Parameters: $K = 16$, $FWHM = 300$ nm, $N = 500$, $SBR = 5$ unless otherwise stated.

In principle, RASTMIN can be performed in any laser-scanning (confocal) microscope, as they are readily prepared to perform rectangular raster scans. The only hardware modification needed would be including a phase-mask into the excitation beam path to produce a focus featuring a central minimum (ideally a zero) of intensity. In this way, the power of localizing with intensity minima could be made available to significantly more optical systems available in many labs. We note, however, that achieving nanometer localization precision requires active stabilization or drift correction systems with nanometer accuracy.

We also analyze the performance of the counterpart of RASTMIN using excitation maxima. For a sufficiently large $L$ this approach is equivalent to conventional laser-scanning (confocal) imaging and localization of the single emitter which has recently been reported by the group of Jörg Enderlein and named Confocal Fluorescence Lifetime SMLM (FL-SMLM)[29]. However, to avoid implying that confocality or picosecond time-resolved detection are necessary conditions for this method, we will name it RASTMAX as a more general approach that would include any technique that raster scans a (Gaussian) maximum of light over a single emitter.

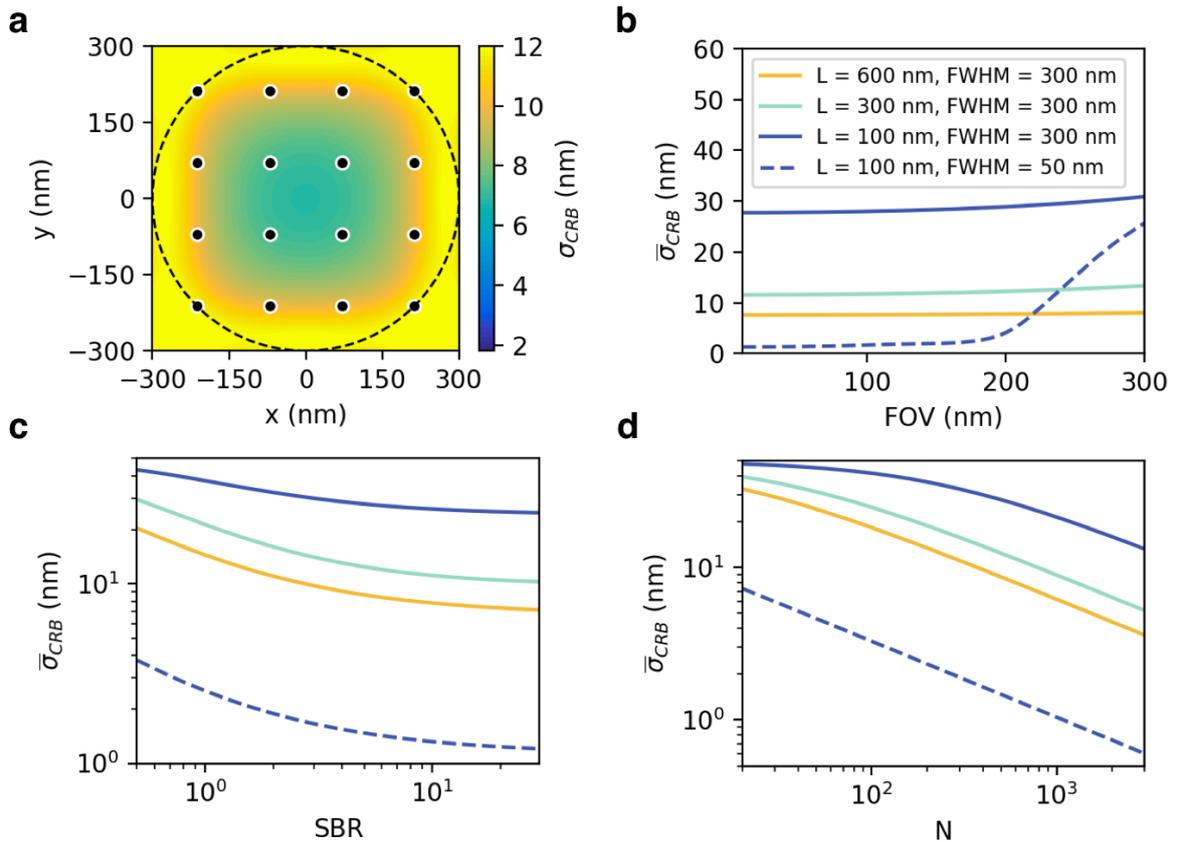

**Figure 6. Raster scanning with a maximum.** (a) Precision map $\sigma_{CRB}(x,y)$ for $L = 600$ nm. Black dotted line indicates a circle of diameter $L$, black dots indicate the positions $r_i$ of the exposures. (b) $\bar{\sigma}_{CRB}$ as a function of the $FOV$. (c) $\bar{\sigma}_{CRB}$ as a function of $SBR$. (d) $\bar{\sigma}_{CRB}$ as a function of $N$. Parameters:

$K = 16, N = 500, SBR = 5$ unless otherwise stated. $FWHM = 300$ nm (solid lines), $FWHM = 50$ nm (dotted line).

Figure 6a shows a map of $\sigma_{CRB}$ for RASTMAX with $FWHM = 300$ nm, $L = 600$ nm, $N = 500$ and $SBR = 5$. Within the region-of-interest defined by $FOV = 0.75 \, L$, the average localization precision ranges from 7 to 9 nm. Contrary to what happens in RASTMIN, excitation patterns smaller than the $FWHM$ of the excitation beam decrease the precision achieved by RASTMAX (Figure 6b, solid lines). Given a certain $FWHM$, we find that $L < FWHM$ gives poor results in terms of precision because the part of the excitation beam with more sensitivity, the flanks of the gaussian focus, are not used to excite the emitter. On the other hand, using $L \gg FWHM$ is not optimal either because most exposures would not excite the emitter efficiently and only contribute to add background to the measurement. Hence an optimal situation is given by $L \sim 2 \, FWHM$.

It is of interest to analyze RASTMAX with sub-diffraction excitation maxima, attained for example through STED. To our knowledge, such a nanoscopy scheme has not yet been realized, although experimental results of STED nanoscopy on immobilized single molecules have been reported[38,39]. We study the potential performance of such a method by considering a RASTMAX scheme with $FWHM = 50$ nm and $L = 100$ nm. As it can be seen in Figure 6b (blue, dotted line) such a method has the potential to reach precisions comparable to MINSTED.

RASTMAX precision as a function of $FOV$ remains fairly constant up to $FOV = 2L$ where it starts to decrease mainly due to a drop in relative $SBR$ (Figure 6b). On the other hand, the precision as a function of $SBR$ decays similarly to the other methods (Figure 6c). The calculations indicate that $\sim 4$ nm precision should be reached for $N \sim 1000$ with a $SBR = 5$ (Figure 6d). While it does not match the precisions of MINFLUX, OTMIN, or RASTMIN, RASTMAX should significantly outperform camera-based SMLM. The reason for this is that the measurement process in a single-photon counting detector such as avalanche photo-diodes is well described by Poisson noise, while detecting with a camera involves other sources of noise that compromise localization precision at relatively low photon numbers[9]. A comparison between RASTMAX and a hypothetical camera detection with purely Poisson noise is described in Supplementary Section 2.

## TOP PERFORMANCE COMPARISON

Finally, we made a comparison of reported and new methods under optimum conditions for each one. Figure 7 summarizes these results. The already known methods (OT, MINSTED, MINFLUX, RASTMAX) were evaluated using the best combinations of parameters that have been experimentally realized. For the new methods (OTMIN, RASTMIN), we chose optimum parameters that are experimentally realizable. A $FWHM = 300$ nm was used for all diffraction-limited foci. When a sub-diffraction maximum of intensity was used (MINSTED) we considered a $FWHM = 50$ nm. Each method was evaluated for the best-performing and realistic value of $L$: $L_{OT} = 300$ nm, $L_{RASTMAX} = 600$ nm, and $L_{MINSTED} = L_{MINFLUX} = L_{RASTMIN} = L_{OTMIN} = 50$ nm. In all cases, $SBR = 5$ and a total photon-count $N = 500$ were considered.

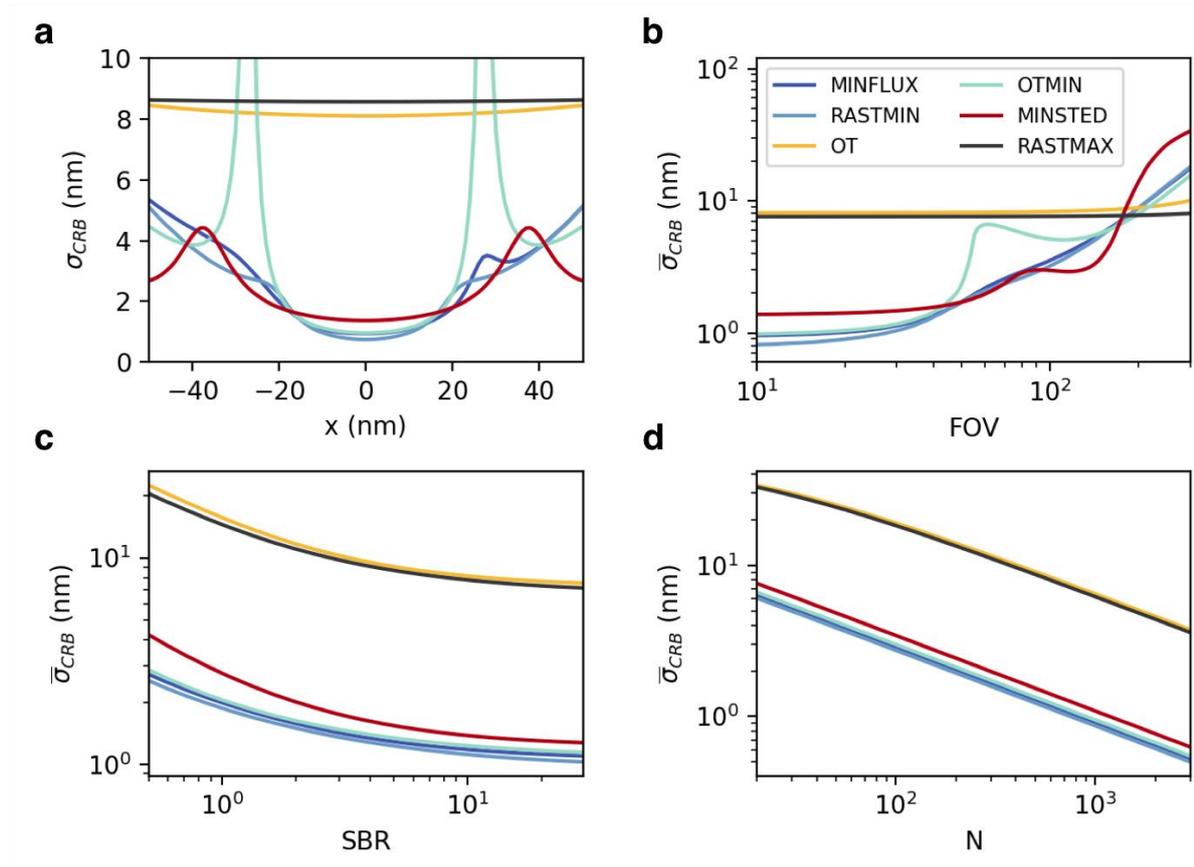

**Figure 7. Comparison of the different methods.** (a) 1D profile ($y = 0$) of the precision map $\sigma_{CRB}(x, y)$ for all the methods using their best-performing realistic parameters. A $FWHM = 300$ nm was used for all diffraction-limited foci. For MINSTED we considered a $FWHM = 50$ nm. $L$: $L_{OT} = 300$ nm, $L_{RASTMAX} = 600$ nm, and $L_{MINSTED} = L_{MINFLUX} = L_{RASTMIN} = L_{OTMIN} = 50$ nm. (b) $\bar{\sigma}_{CRB}$

as a function of the $FOV$. (c) $\bar{\sigma}_{CRB}$ as a function of $SBR$. (d) $\bar{\sigma}_{CRB}$ as a function of $N$. Other parameters: $N = 500$, $SBR = 5$ unless otherwise stated.

Among the methods that use diffraction-limited excitation, the ones using a minimum of intensity achieve a ~5 fold better precision than the ones using a maximum, regardless of the sequence of exposures (Figure 7a). Methods using sub-diffraction excitation maxima (i.e. MINSTED) can achieve a precision up to ~1 nm by engineering an effective PSF well below the diffraction limit of light.

In general, all techniques present the best performance in the central region of the excitation pattern, over an area about 75-80% of the range defined by the $L$ (Figure 7b). In this regard, methods featuring maxima are more robust and perform well over larger regions of space.

For $SBR > 5$, the localization precision of all techniques is always better than 75% of the ideal precision for infinite $SBR$ (Figure 7c). Detailed numbers on the analysis of precision with respect to $SBR$ and $FOV$ are given in Supplementary Table 1. Methods using a minimum of intensity are $\sim 10 - 20$ times more photon efficient, reaching molecular-scale precision ($\sigma_{CRB} \sim 1$ nm) with $N \sim 1000$. Methods using a maximum of intensity are limited to $\sigma_{CRB} \sim 3 - 5$ nm for $N = 1000 - 3000$ and require much higher photon budgets ($N \geq 30000$) to achieve $\sigma_{CRB} \sim 1$ nm.

## CONCLUSIONS AND OUTLOOK

We have presented a framework that is common to all single-molecule localization techniques that use a sequence of excitations with structured illumination. Under this framework, based on information theory and maximum likelihood estimation, we made a fair comparison between methods using the Cramér-Rao bound, which is independent of the estimator used to infer the position of the emitter. Only the Poisson shot-noise of the photon counts was considered. In this way, we computed the maximum possible localization precision, which is attainable with modern single-photon counting detectors such as avalanche photodiodes. Naturally, the analysis could be extended to represent other detectors by including additional sources of noise. Another advantage of the mathematical framework is the possibility to include formally the prior information needed in these methods to pre-locate the molecules. While we have focused on 2D localization, it is straightforward to generalize the analysis to three dimensions.

The common framework makes it easy to design new approaches. Here, we presented two new single-molecule localization schemes: OTMIN and RASTMIN. Both schemes achieve the highest localization precision, similar to MINFLUX, and have the potential to be implemented in existing optical systems with minor changes. OTMIN could be implemented in any OT setup by just adding a suitable phase mask to engineer a light focus with a minimum. A similar approach can be used to implement RASTMIN in any laser-scanning (confocal) microscope. We believe that these two approaches, and RASTMIN in particular, can significantly contribute to a wider application of fluorescence nanoscopy with molecular-scale resolution.

We found that all approaches featuring an intensity minimum have a similar performance in the central region of the excitation pattern. Independently of the geometry of the excitation pattern, they outperform methods featuring an intensity maximum by at least a factor of 5, reaching molecular-scale precision ($\sim 1$ nm) with only $N \sim 1000$ detected photons at a $SBR = 5$.

In practice, RASTMAX (or confocal-SMLM) provides a significant improvement over camera-based SMLM in terms of precision. This is due to the fact that photon detection with current avalanche photodiodes includes almost only Poisson noise, while EM-CCD or sCMOS cameras present substantial additional noise.

All of these methods could benefit from iterative and adaptive approaches that update the sequence of excitations with new information about the position of the emitter, as it was done with MINFLUX[11]. Moreover, while confocal detection is not necessary, it could be advantageous to obtain higher $SBR$ conditions. We also note that active $xyz$ drift compensation could be key to attain the highest localization precisions.

Methods that use sub-diffraction effective excitation patterns such as MINSTED or a combination of RASTMAX and STED can achieve localization precisions as good as methods using minima of light. However, it should be mentioned that in these experiments the total number of detected fluorescence photons $N$ usually corresponds to a much higher number of excitation-emission cycles than in conventional measurements, with the consequent stress on the photostability of the emitter.

Finally, we note that other position estimators might be more suitable than MLE for different reasons (computational efficiency for real-time calculations, unbiased estimators at low $N$, etc). However, we believe that our approach explains thoroughly the fundamental similarities and differences between the different existing methods and will also be a powerful tool to design, develop, and combine new single-molecule localization methods and experiments.

**FUNDING**

This work has been funded by CONICET, ANPCYT Projects PICT-2013-0792, and PICT-2014-0739.

**ACKNOWLEDGMENTS**

The authors acknowledge Mariela Sued and Daniela Rodríguez for discussions on the interpretation of some of the results of this work. L.A.M thanks Iván Lengyel and Felipe Marceca for help with technical aspects of this work and Andrew G. York for inspiring conversations on inverse problems. F.D.S. acknowledges the support of the Max-Planck-Society and the Alexander von Humboldt Foundation.

**AUTHOR CONTRIBUTIONS**

L.A.M and F.D.S. conceived the main conceptual ideas of this work. L.A.M developed the mathematical framework and performed all the calculations and simulations presented in this work. L.F.L performed calculations and simulations at an early stage of the project. F.D.S supervised the project. All authors discussed the results and wrote the manuscript.

**DECLARATION OF INTERESTS**

The authors declare no competing financial interest.

Supporting Information for

# A common framework for single-molecule localization using sequential structured illumination

Luciano A. Masullo[1,2], Lucía F. Lopez[1], Fernando D. Stefani[1,2]

1 Centro de Investigaciones en Bionanociencias (CIBION), Consejo Nacional de Investigaciones Científicas y Técnicas (CONICET), Godoy Cruz 2390, C1425FQD, Ciudad Autónoma de Buenos Aires, Argentina

2 Departamento de Física, Facultad de Ciencias Exactas y Naturales, Universidad de Buenos Aires, Güiraldes 2620, C1428EHA, Ciudad Autónoma de Buenos Aires, Argentina

## Table of contents

# Supplementary Section 1. Derivation of $p_i(r_E)$ with background

If we assume pure-Poisson noise and we set the number of detected photons to $N$, the resulting measured array of photons has a multinomial distribution $n_i \sim multinomial(p_i, N)$. In the absence of background the multinomial parameters $p_i$ are given by

$$p_i(r_E) = \frac{\lambda_i(r_E)}{\sum_{j=1}^{K} \lambda_j(r_E)} = \frac{I(r_E - r_i)}{\sum_{j=1}^{K} I(r_E - r_j)} \tag{S1}$$

That is the ratio between the excitation intensity of the $i$-th exposure and the sum of the rest of the exposures. Here, a linear relationship between the expected detected photon counts in each exposure $\lambda_i$ and the intensity $I(r_E - r_i)$ is assumed. This is a very good approximation for fluorescence microscopy within the linear regime (far from saturation). In the case of methods using other photophysical transitions, the linear relationship will still hold with the *effective* excitation intensity $I_{eff}(r_E - r_i)$.

In the presence of background equation (S1) takes the form

$$p_i(r_E) = \frac{\lambda_i(r_E) + \lambda_{b_i}(r_E)}{\sum_{j=1}^{K} \lambda_j(r_E) + \lambda_{b_j}(r_E)} \tag{S2}$$

Following standard definitions, we can now define a signal-to-background ratio function

$$SBR(r_E) \equiv \frac{\sum_{j=1}^{K} \lambda_j(r_E)}{\sum_{j=1}^{K} \lambda_b(r_E)} = \frac{\sum_{j=1}^{K} \lambda_j(r_E)}{K\,\lambda_b} \tag{S3}$$

where we have assumed $\lambda_{b_i}(\mathbf{r_E}) = \lambda_b \ \forall i$, that is that the detected background does not depend on the position of the single emitter $\mathbf{r_E}$ and that all the background contributions of each exposure are approximately equal. This is a very good approximation of an experimental situation in which most background will come from out-of-focus autofluorescence coming from a biological context or the coverslip or other optical components. Also, the diffraction-limited size of the detection volume is considerably larger than the usually sub-diffraction excitation pattern. Thus, the background generated by the excitation field $I(\mathbf{r} - \mathbf{r_i})$ is practically independent of the position $\mathbf{r_i}$.

Hence, using that $SBR\ K\ \lambda_b = \sum_{j=1}^{K} \lambda_j(\mathbf{r_E})$, we can rewrite

$$p_i(\mathbf{r_E}) = \frac{\lambda_i(\mathbf{r_E}) + \lambda_b}{SBR\ K\lambda_b + K\lambda_b} = \frac{\lambda_i(\mathbf{r_E}) + \lambda_b}{K\lambda_b(SBR + 1)} \tag{S4}$$

Multiplying and dividing by $\sum_{j=1}^{K} \lambda_j(\mathbf{r_e})$ and using the definition (S3) we obtain

$$p_i(\mathbf{r_E}) = \frac{\lambda_i(\mathbf{r_E}) + \lambda_b}{K\lambda_b\ (SBR + 1)} \frac{\sum_{j=1}^{K} \lambda_j(\mathbf{r_E})}{\sum_{j=1}^{K} \lambda_j(\mathbf{r_E})} = \frac{\lambda_i(\mathbf{r_E})}{\sum_{j=1}^{K} \lambda_j(\mathbf{r_E})} \frac{SBR}{(SBR + 1)} + \frac{1}{K(SBR + 1)} \tag{S5}$$

And hence,

$$p_i(\mathbf{r_E}) = \frac{SBR(\mathbf{r_E})}{SBR(\mathbf{r_E}) + 1} \frac{I(\mathbf{r_E} - \mathbf{r_i})}{\sum_{j=1}^{K} I(\mathbf{r_E} - \mathbf{r_j})} + \frac{1}{SBR(\mathbf{r_E}) + 1} \frac{1}{K} \tag{S6}$$

where we explicitly write the dependence of $SBR$ with the position of the emitter.

# Supplementary Section 2. RASTMAX (confocal scan) vs image-based localization with a camera

When comparing these two methods, it is important to note that they are based on two different physical phenomena. Single-molecule localization by sequential structured illumination obtains the molecular position information from light absorption. The differences in molecular excitation at each exposure of the sequence lead to different fluorescence emission intensities that are detected with a single photodetector. By contrast, in camera-based single-molecule localization, illumination is uniform and all the information about the molecular position is obtained from the angular photon emission registered as an image in an array of photodetectors (camera). Despite this, there are similarities in their position estimation and performance.

In camera-based single-molecule localization, emitted photons are detected in each camera pixel with a certain probability related to the image intensity at that pixel, which can be approximated by a Gaussian function, in this case corresponding to the point-spread function of the optical system. Hence, following an analog procedure to the one described in Supplementary Section 1, we can write:

$$p_i(\boldsymbol{r_E}) = \frac{SBR(\boldsymbol{r_E})}{SBR(\boldsymbol{r_E})+1} \frac{Int\_Gauss\,(\boldsymbol{r_E}-\boldsymbol{r_i})}{\sum_{j=1}^{K} Int\_Gauss(\boldsymbol{r_E}-\boldsymbol{r_j})} + \frac{1}{SBR(\boldsymbol{r_E})+1} \frac{1}{K} \quad (S7)$$

Where $\boldsymbol{r_i} = (x_i, y_i)$ defines the central position of the $i$-th pixel of the camera and $\boldsymbol{r_E} = (x_E, y_E)$ is the position of the emitter. $Int\_Gauss\,(\boldsymbol{r_E}-\boldsymbol{r_i})$ is the integral of the Gaussian image intensity over the area of the $i$ pixel:

$$Int\_Gauss(\boldsymbol{r_E}-\boldsymbol{r_i}) = \int_{y_i-\frac{a}{2}}^{y_i+\frac{a}{2}} \int_{x_i-\frac{a}{2}}^{x_i+\frac{a}{2}} Gauss(x-x_E, y-y_E)dxdy \quad (S8)$$

With

$$Gauss(x - x_E, y - y_E) = A_{Gauss} \, exp\left(-\frac{1}{2}\frac{(x-x_E)^2}{\sigma_{PSF}^2}\right) exp\left(-\frac{1}{2}\frac{(y-y_E)^2}{\sigma_{PSF}^2}\right) \quad (S9)$$

where $a$ is the pixel width and height (assumed to be squared), $\sigma_{PSF}$ defines the size of the Gaussian $PSF$ (assumed to be symmetrical) related to the $FWHM$ by $FWHM \approx 2.35 \, \sigma_{PSF}$, and $A_{Gauss}$ is an amplitude that will cancel out when computing $p_i(\boldsymbol{r_E})$. The expected background contribution $\lambda_b$ is again assumed to be constant, equal for all pixels.

The analogy between RASTMAX and a camera-based approach becomes evident. The integral in equation (S8) can be approximated by $a^2 Gauss(x - x_E, y - y_E)$. Then, equation (S7) becomes:

$$p_i(\boldsymbol{r_E}) = \frac{SBR(\boldsymbol{r_E})}{SBR(\boldsymbol{r_E}) + 1} \frac{Gauss\,(\boldsymbol{r_E} - \boldsymbol{r_i})}{\sum_{j=1}^{K} Gauss(\boldsymbol{r_E} - \boldsymbol{r_j})} + \frac{1}{SBR(\boldsymbol{r_E}) + 1}\frac{1}{K} \quad (S10)$$

Which is formally equal to equation (5) of the manuscript with $I = I_{Gauss}$. In both cases the distribution of detected photons is Gaussian. In one case due to a Gaussian illumination and the other due to a Gaussian image.

Supplementary Figure 2 shows example simulations of camera-based SMLM using equation (S7) and RASTMAX for $N = 500$ and $SBR = 5$. The difference in wavelength due to the expected Stokes shift was neglected and the same $FWHM = 300$ nm was used in both cases. As expected, the localization precision attained by both methods is practically identical.

In practice, camera-based approaches cannot reach this level of precision. Reported values are typically a factor of 2 to 3 worse[1,2]. The reason for this is that cameras present other sources of noise in addition to the fundamental Poisson shot-noise.

# Supplementary Figure 1

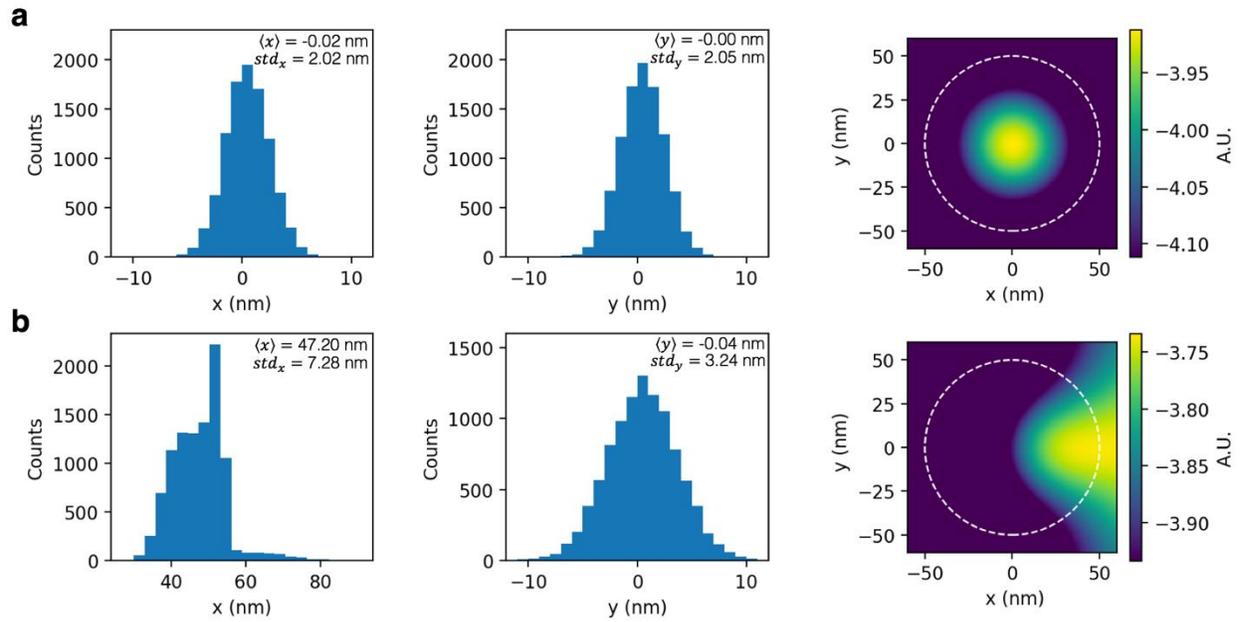

**Supplementary Figure 1. Simulated localizations for OTMIN.** Estimations of $x$, estimation of $y$, and average likelihood function using OTMIN with $L = 100$ nm for **(a)** $r_E = (0, 0)$ and **(b)** $r_E = (50, 0)$. CRB values for **(a)** $\sigma_{CRB_x} = 2.01$ nm, $\sigma_{CRB_y} = 2.01$ nm. Parameters: $N = 500$, $SBR = 5$. Simulation size: 10000 samples for each position.

## Supplementary Figure 2

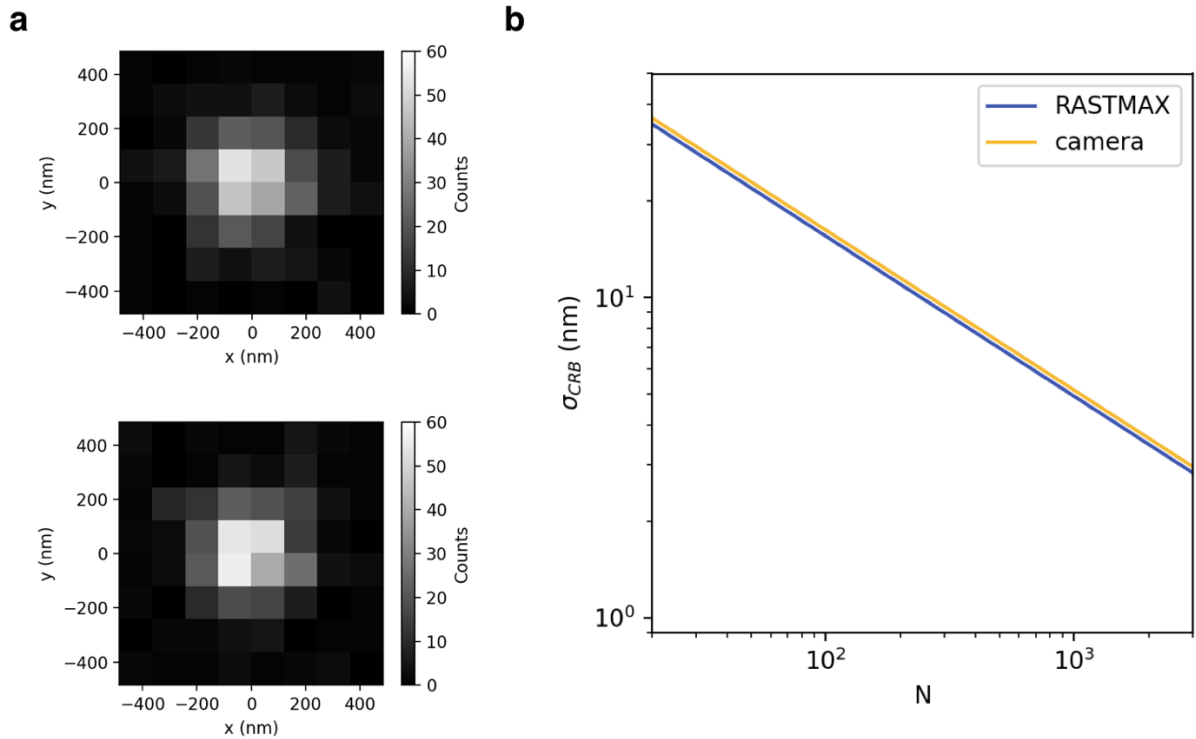

**Supplementary Figure 2. Comparison between RASTMAX and camera-based SM localization**. (a) Simulation of single-molecule localization experiment in a widefield excitation, camera-based detection (top), and a raster scanning with a gaussian beam in excitation and a single detector (bottom). Parameters: $N = 500$, $SBR = 5$. (b) $\sigma_{CRB}(0,0)$ as a function of $N$, for RASTMAX and camera-based localization. $SBR = 5$. No *prior* was used in the calculated CRB for each method.

# Supplementary Table 1

| CONFIGURATION | $\sigma_{CRB}(r_E = 0)$ | $\bar{\sigma}_{CRB}$ | $SBR = 10000$ | $SBR = 5$ |
|---|---|---|---|---|
| MINFLUX L 50 | 0.94 | 1.29 | 1.05 | 1.29 |
| MINFLUX L 100 | 1.96 | 2.65 | 2.16 | 2.65 |
| MINFLUX L 150 | 3.16 | 4.18 | 3.39 | 4.18 |
| OTMIN L 50 | 0.96 | 1.34 | 1.09 | 1.34 |
| OTMIN L 100 | 2.00 | 2.77 | 2.22 | 2.77 |
| OTMIN L 150 | 3.23 | 4.36 | 3.43 | 4.36 |
| OTMIN L 50 K 6 | 0.96 | 1.34 | 1.10 | 1.34 |
| OTMIN L 100 K 6 | 2.00 | 2.78 | 2.28 | 2.78 |
| OTMIN L 150 K 6 | 3.23 | 4.38 | 3.51 | 4.38 |
| RASTMIN L 50 | 0.74 | 1.23 | 0.97 | 1.23 |
| RASTMIN L 100 | 1.52 | 2.47 | 1.93 | 2.47 |
| RASTMIN L 150 | 2.35 | 3.71 | 2.88 | 3.71 |
| RASTMAX L 100 | 27.73 | 27.87 | 24.26 | 27.86 |
| RASTMAX L 300 | 11.56 | 12.44 | 9.80 | 12.44 |
| RASTMAX L 600 | 7.59 | 8.69 | 6.72 | 8.69 |
| RASTMAX L 100, FWHM 50 | 1.28 | 1.47 | 1.13 | 1.47 |
| OT L 50 (MINSTED) | 1.37 | 1.54 | 1.22 | 1.54 |
| OT L 100 (MINSTED) | 2.73 | 3.07 | 2.43 | 3.07 |
| OT L 300 | 8.1 | 9.07 | 7.23 | 9.07 |
| OT L 50 K 6 (SMCT) | 1.37 | 1.54 | 1.22 | 1.54 |
| OT L 100 K 6 (SMCT) | 2.73 | 3.07 | 2.43 | 3.07 |
| OT L 300 K 6 (SMCT) | 8.1 | 9.07 | 7.23 | 9.07 |

**Supplementary Table 1**. Performance comparison for various schemes of SML-SSI- Column 1: SML-SSI method and parameter $L$. Column 2: localization precision at the center of the excitation pattern $\sigma(r_E = (0,0))$. Column 3: average precision $\bar{\sigma}_{CRB}$ over a FOV corresponding to a circular area of diameter $0.75\ L$ at $SBR = 5$ and $N = 500$. Column 4: average precision $\bar{\sigma}_{CRB}$ at $SBR = 10000$ (considered as infinite) and $N = 500$. Column 5: average precision $\bar{\sigma}_{CRB}$ at $SBR = 5$ and $N = 500$. All values are in nm.

# Suppementary References